\documentclass[aps,showpacs,amsmath,amssymb,floatfix,twocolumn]{revtex4}

\usepackage{graphicx}

\usepackage{epsfig}

\usepackage{amssymb,amsmath,amsfonts}

\newcommand{\ga}{\alpha}

\newcommand{\gb}{\beta}

\newcommand{\be}{{\begin{equation}}}

\newcommand{\ee}{{\end{equation}}}

\newcommand{\ba}{{\begin{eqnarray}}}

\newcommand{\ea}{{\end{eqnarray}}}

\begin{document}

\title{QCD equation of state in a virial expansion}

\author{S.~Mattiello and W.~Cassing}

\affiliation{Institute for Theoretical Physics, University of Gie{\ss}en, Germany}

\begin{abstract}

We describe recent three-flavor QCD lattice data for the pressure, speed of sound and interaction measure at nonzero temperature and vanishing chemical potential within a virial expansion.
For the deconfined phase we use a phenomenological model which includes non-perturbative effects from dimension two gluon condensates that reproduce the free energy of quenched QCD very well.
The hadronic phase is parameterized by a generalized resonance-gas model.
Furthermore, we extend this approach to finite quark densities introducing an explicit $\mu$-dependence of the interaction. We calculate pressure, quark-number density, entropy and energy density and compare to results of lattice calculations.
We, additionally, investigate the structure of the phase diagram by calculating the
isobaric and isentropic lines as well as the critical endpoint in the ($T, \mu_q$)-plane.
\end{abstract}

\maketitle


\section{Introduction}\label{Sec:introduction}

The exploration of the phase structure of quantum chromodynamics (QCD) for the
whole density-temperature plane is a challenging task for the standard
model.
Lattice Monte Carlo simulations have revealed several exciting results over the past decade~\cite{Tannenbaum:2007dx,Fodor:2002km,Fodor:2001pe,Allton:2003vx,Allton:2005gk,D'Elia:2002gd,D'Elia:2004at,Cheng:2007jq}. At finite densities, modeling of QCD has added substantially to our understanding of its rich phase structure~\cite{Alford:1997zt,Alford:1998mk,Rapp:1997zu,Rajagopal:2000wf,Klevansky:1992qe,Bluhm:2004xn,Bluhm:2007nu,Ratti:2005jh,Cassing:2007yg,Cassing:2007nb,Cassing:2008nn,Mattiello:2004rd,Strauss:2009uj}.
In the last years heavy-ion collision experiments focus on the investigation the quark-gluon plasma (QGP).
Early concepts of the QGP were guided by the assumption of a weakly interacting systems of partons because the entropy density $s$ and the energy density $\varepsilon$ calculated from lattice QCD were close to the Stefan Boltzmann (SB) limit for a relativistic noninteracting system. However, recent observations at the Relativistic Heavy Ion Collider (RHIC) indicated that the QGP created in ultrarelativistic Au + Au collision was interacting more strongly than hadronic matter~\cite{Gyulassy:2004zy,Jacobs:2004qv,Stoecker:2004qu,Adcox:2004mh,Shuryak:2004cy,Heinz:2004pj,Song:2007fn}.
Because of these experimental facts the noticeable deviation ($\approx 10-15\%$) became more important.

In the hadronic phase of QCD the partition function of a hadron resonance-gas yields a satisfactory description of lattice results on bulk thermodynamic observables in the low temperature region for $\mu_{\rm q}=0$~\cite{Karsch:2003vd} and in the general case at finite densities, i.e. $\mu_{\rm q}\neq 0$~\cite{Karsch:2003zq}.
In this context the investigation of the interaction between the partons in the deconfined phase plays a crucial role to understand the properties of the QGP.
The dynamical quasiparticle picture is one important way to relax the approximation of the QGP as an ideal gas~\cite{Cassing:2007yg,Cassing:2007nb,Peshier:2004bv,Peshier:2005pp} of massive but noninteracting degrees of freedom.
Alternatively, the phase structure can be investigated in the finite width model (FWM) by analizing its isobaric partition given by a generalization of statistical ensembles with fluctuating extensive quantities~\cite{Bugaev:2007pd,Bugaev:2008jj}. Another possibility is to use a generalized formalism to calculate the corrections to a single-particle partition function starting from an interaction potential.

In this work we use the latter formalism, which is a generalized
version of the classical virial expansion, to calculate the
partition function with an interaction potential extracted from
lattice calculations.\\

The paper is organized as follows: In Section~\ref{Sec:Virial} we
introduce the virial expansion for vanishing quark chemical
potential $\mu_q$. In Section~\ref{Sec:Model} we motivate our
model for the hadronic as well as for the deconfined phase and in
Section~\ref{Sec:Results} we show the results of the virial
expansion within our model. We extend in Section~\ref{Sec:mu} the
virial expansion to finite densities and again compare our results
with lattice calculations. Furthermore, we explicitly evaluate the
confinement-deconfinement phase transition and the isobaric and
isentropic lines. The conclusions in Section~\ref{Sec:Concl}
finalize this work.

\section{Virial expansion at $\mu_q=0$}\label{Sec:Virial}

The main quantity to establish thermodynamical properties of QCD
is the partition function $Z(T,V)$. In this Section we restrict
ourselves to the case of vanishing chemical potential $\mu_q$,
i.e. vanishing baryon number. The partition function then is
obtained as
\begin{equation}
Z(T,V)={\rm Tr}\left[{\rm e}^{-\beta H}\right],
\end{equation}
where $H$ is the Hamiltonian of the system and $\gb=1/T$ is the
inverse temperature. For particles with spin-isospin degeneracy
factor $d$ the logarithm of the partition function reads
\begin{equation}\label{eqn:lnZ}
\ln{Z}=d\int_V\!d^3 {\bf r}\;\int\!d^3 {\bf p}\;\eta\log{\left(1+\eta{\rm e}^{-\gb H({\bf r},{\bf p})}\right)},
\end{equation}
where $\eta=-1$ for bosons and  $\eta=1$ for fermions.

In the case of an interaction free system, the Hamiltonian $H({\bf r},{\bf p})$ is given by the kinetic energy $E_k$ and therefore depends on $p=\vert{\bf p}\vert$ only.
For massless particles one has (with $\hbar=c=1$) $H(p)=p$ and can calculate the partition function for bosons as well as for fermions,
\begin{eqnarray}
\ln{Z_{\rm B}}&=&d_{\rm B}V\frac{\pi^2}{90\gb^3}\equiv V\ln{Z^{(0)}_{\rm B}}\\
\ln{Z_{\rm F}}&=&d_{\rm F}V\frac{7\pi^2}{720\gb^3}\equiv V\ln{Z^{(0)}_{\rm F}}.
\end{eqnarray}
Accordingly, the boson and the fermion partition functions are
connected by
\begin{equation}
d_{\rm B}\ln{Z^{(0)}_{\rm F}}=\frac{7}{8}d_{\rm F}\ln{Z^{(0)}_{\rm B}}.
\end{equation}
For non-interacting particles with mass $m$ the Hamiltonian is
given by $H(p)=\sqrt{p^2+m^2}$ and the partition function can be
expressed in terms of the Bessel function $K_2$.\\ From the
partition function  all thermodynamic quantities can be derived:
pressure, energy density, entropy, interaction measure and
velocity of sound. Using the free massless partition function this
leads to the well-known Stefan Boltzmann limit.\\ To relax the
dependence on the size of the system we divide the partition
function by the volume $V$ and all calculated quantities become
densities. In general we are not able to calculate exactly the
partition function except by numerical diagonalization of the
many-body Hamiltonian. To incorporate the effect of interactions
we start with the classical virial expansion. To this aim we
consider the $N$-body Hamiltonian defined by
\begin{eqnarray}
H_1({\bf r_1},{\bf p_1})&=&E_ {\rm k}(p_1)\\
H_2({\bf r_1},{\bf r_2},{\bf p_1},{\bf p_2})&=&E_ {\rm k}(p_1)+E_ {\rm k}(p_2)+W_{12}\\
H_3({\bf r_1},{\bf r_2},{\bf r_3},{\bf p_1},{\bf p_2},{\bf p_3})&=&E_ {\rm k}(p_1)+E_ {\rm k}(p_2)\nonumber\\
&&+E_ {\rm k}(p_3)+W_{123}\\
\cdots\nonumber&&
\end{eqnarray}
In classical physics the Bose as well the Fermi distributions in Eq. (\ref{eqn:lnZ}) merge to the Boltzmann distribution and the logarithm in the integrand drops.
Then, we can separate the kinetic and the interaction term in the partition function.
To recover the quantum statistics we perform a minimal substitution by replacing the integral over the momentum by the integral for the free case, which leads to the exact Stefan Boltzmann limit.
The definition
\begin{equation}
\zeta=\ln{Z^{(0)}},
\end{equation}
where we suppressed for simplicity the indices for fermion and bosons,
leads to an expansion of the partition function in powers of $\zeta$, i.e.
\begin{equation}
Z=\sum_{\nu=0}^{\infty}\frac{b_\nu}{\nu!}\zeta^\nu.
\end{equation}
We consider only the first two terms of this expansion:
then the coefficients are given by
\begin{eqnarray}
b_1&=&1\\
b_2&=&\int_V\!d^3 {\bf r}\;\left({\rm e}^{-\gb W_{12}(r)}-1\right).\label{eqn:b_2}
\end{eqnarray}
The pressure reads
\begin{equation}\label{eqn:P}
P=T\ln{Z}=T\sum_{\nu=1}^{\infty}\frac{b_\nu}{\nu!}\zeta^\nu,
\end{equation}
and can be directly expressed in terms of the particle density $\rho$.
We obtain for the first terms of the expansion
\begin{equation}\label{eqn:Pdens}
P=T\rho\left(1-\frac{b_2}{2}\rho+\cdots\right).
\end{equation}
The other quantities can be calculated from the pressure
or the partition function using the thermodynamic relations:
\begin{eqnarray}
\varepsilon&=&-\frac{\partial\ln{Z}}{\partial\beta}\label{eqn:eps}\\
s&=&\frac{\partial P}{\partial T}\label{eqn:S}.\\
c^2_{\rm s}&=&\frac{dP}{d\varepsilon}
=\varepsilon\frac{dP/\varepsilon}{d\varepsilon}+\frac{P}{\varepsilon} \label{eqn:c^2_s}.
\end{eqnarray}

\section{The Model}\label{Sec:Model}

The application of the virial expansion formalism to investigate
the properties of the hadronic and partonic matter requires the
calculation of the single-particle partition function for the
degrees of freedom and of the corrections given by the residual
interaction between the 'constituents'. For the confined phase the
degrees of freedom are the hadrons while in the QGP phase they are
the quarks/antiquarks as well as the gluons.

The analysis of thermodynamic properties of the hadronic phase has
demonstrated that a good description is achieved using a
phenomenological resonance-gas model taking into account all
mesonic and baryonic resonances with masses up to 1.8 GeV and 2.0
GeV, respectively. This amounts to 1026 resonances that truly are
experimental parameters of the model~\cite{Cheng:2007jq}. In this
case the residual interaction may be neglected. However, to reduce
the number of the experimental parameters for the description of
the low temperature region of QCD we explicitly write the
partition function of the resonance-gas in the Boltzmann limit
as~\cite{Blanchard:2004du}
\begin{equation}\label{eqn:Z_HGS}
\ln{Z(V,T)}=\sum_{i=1}\frac{VTm^2_i}{2\pi^2}\rho(m_i)K_2\left(\frac{m_i}{T}\right),
\end{equation}
where the sum begins with the stable ground state $m_0$ and
includes the resonances $m_i, i=1,2,\cdots$ with weights
$\rho(m_i)$ relative to $m_0$. To specify $\rho(m_i)$ we employ
the Hagedorn hypothesis~\cite{Hagedorn:1965st,Broniowski:2004yh}
of an exponential growth of the number of hadronic resonances with
mass. The formula for the asymptotic dependence of the density of
hadronic states on mass $m$ is given by
\begin{equation}\label{def:rho}
\rho(m)=f(m)\exp{\{m/T_{\rm H}\}},
\end{equation}
where $f(m)$ denotes a slowly varying function and $T_{\rm H}$ is
the Hagedorn temperature estimated to be in the range of $180-210$
MeV. We choose the form
\begin{equation} \label{form1}
f(m)=am^{-\ga},\quad \ga=3/2
\end{equation}
defined for $m_0\geq 140$ MeV taking into account that the pions
are the hadronic states with the lowest mass. We now replace the
sum in the resonance-gas partition function (\ref{eqn:Z_HGS}) by
an integral and insert the exponentially growing mass spectrum
(\ref{def:rho}) with the choice of $f(m)$ (\ref{form1}),
\begin{equation}\label{eqn:Z_HGI}
\ln{Z(V,T)}=\frac{aVT}{2\pi^2}\int_{m_0}^{\infty}dm \
\!\sqrt{m}\exp{\{m/T_{\rm H}\}}K_2\left(\frac{m}{T}\right).
\end{equation}
In the limit of low temperatures, i.e. $z=m/T\gg 1$, the Bessel
function can be approximated by,
\begin{equation}\label{def:K_2}
K_2(z)=\sqrt{\frac{\pi}{2z}}\exp{\{-z\}},
\end{equation}
and the integral in the resonance-gas partition function
(\ref{eqn:Z_HGI}) can be solved analytically. This gives
\begin{equation}\label{eqn:Z_HG}
\ln{Z(V,T)}=aV\left(\frac{T}{2\pi}\right)^{3/2}\frac{\exp{\{-m_0b\}}}{b}, \quad b=\frac{1}{T}-\frac{1}{T_{\rm H}},
\end{equation}
and leads to a good description of the thermodynamical variables.

To perform a realistic investigation of the properties of the QGP
we have to determine a realistic effective interaction between the
partons in the deconfined phase. The starting point for this aim
is provided by lattice calculations for the quark-antiquark
free-energy. There are many parametrizations of this quantity and
several discussions about the role as effective potentials (or
alternatively as the internal energy)~\cite{Liao:2008wb}. We adopt
the parametrization of the free energy given by Megias et al. in
Ref. ~\cite{Megias:2007pq}. This phenomenological model includes
non-perturbative effects from dimension two gluon condensates.
This one-gluon exchange model, introduced in
Ref.~\cite{Megias:2005ve} for the description of the Polyakov
loop, allows for a good parametrization of the singlet heavy
quark-antiquark free energy lattice data in gluon-dynamics. The
connection between dimension two gluon condensates and
QGP-properties has been confirmed by recent fits of the
interaction measure in gluon-dynamics in terms of the
condensate~\cite{Megias:2008rm,Megias:2008dv}.

The expression for the free energy in this model is given by
\begin{equation}
F_1(r,T)=V_1(r,T)+F_1(\infty,T),
\end{equation}
where $F_1(\infty,T)$ is the singlet free energy at infinite separation and $V_1$ is the effective potential for the deconfined phase.
We use
\begin{equation}
V_1(r,T)=-\left(\frac{\pi}{12}\frac{1}{r}+\frac{{\mathcal C}_2}{2N_{\rm c}T}\right){\rm e}^{-M(T)r},
\end{equation}
where ${\mathcal C}_2$ is the non-perturbative dimension two condensate and $M(T)$ a
Debye mass estimated as
\begin{equation}
M(T)=\sqrt{N_{\rm c}/3+N_{\rm f}/6}\; gT=\tilde g T.
\end{equation}
The parameters have been determined in Ref.~\cite{Megias:2007pq}
from a fit to the lattice data of the Polyakov loop in
gluon-dynamics, i.e. $N_{\rm f}=0$,
\begin{eqnarray}
{\mathcal C}_2&=&(0.9 {\rm GeV})^2\label{eqn:C2fitMegias}\\
\tilde g_{\rm PL}&=&1.26.\label{eqn:gfitMegias}
\end{eqnarray}
In our calculations we will consider light quarks instead of heavy
quarks (with infinite mass) and hence expect deviations from the
previous values  (\ref{eqn:C2fitMegias}) and
(\ref{eqn:gfitMegias}).
Accordingly, we consider $\tilde g$ as a free parameter that we
may fix in comparison with $N_{\rm f }=3$ lattice
calculations~\cite{Cheng:2007jq} for the pressure (calculated for
$N_\tau=6$). These lattice calculations imply a critical
temperature $T_{\rm c} \approx 196$ MeV for the deconfinement phase
transition.

\section{Results}\label{Sec:Results}

Using the one-gluon exchange potential we can directly calculate
the pressure from Eq. (\ref{eqn:P}) for the QGP. To calculate the
second virial coefficient $b_2$ we linearize the exponential
function in  Eq.(\ref{eqn:b_2}). For the hadronic phase we
calculate the pressure from the partition function using Eq.
(\ref{eqn:P}). By comparing the lattice results of the pressure
$P/T^4$ as a function of the temperature for the deconfined phase
(given by $T\geq T_{\rm c}=196$ MeV) with our results we find a
good agreement with $\tilde g$ in the range $1.25 \leq \tilde g\leq 1.35$.
Note that these values for $\tilde g$ are comparable
with the result from the fit of the Polyakov loop,
i.e. $\tilde g_{\rm PL}=1.26$. Our fit in the deconfined region yields
$\tilde g=1.30$ and we can fix the value of the Hagedorn temperature
$T_{\rm H}$ and the constant $a$ to obtain a continuous function
for all temperatures $T>0$. This leads to $T_{\rm H}=205$ MeV and
$a=9.17$ ${\rm  MeV}^{1/2}$.
\begin{figure}[b]
\begin{minipage}[b]{0.45\textwidth}
\epsfig{figure=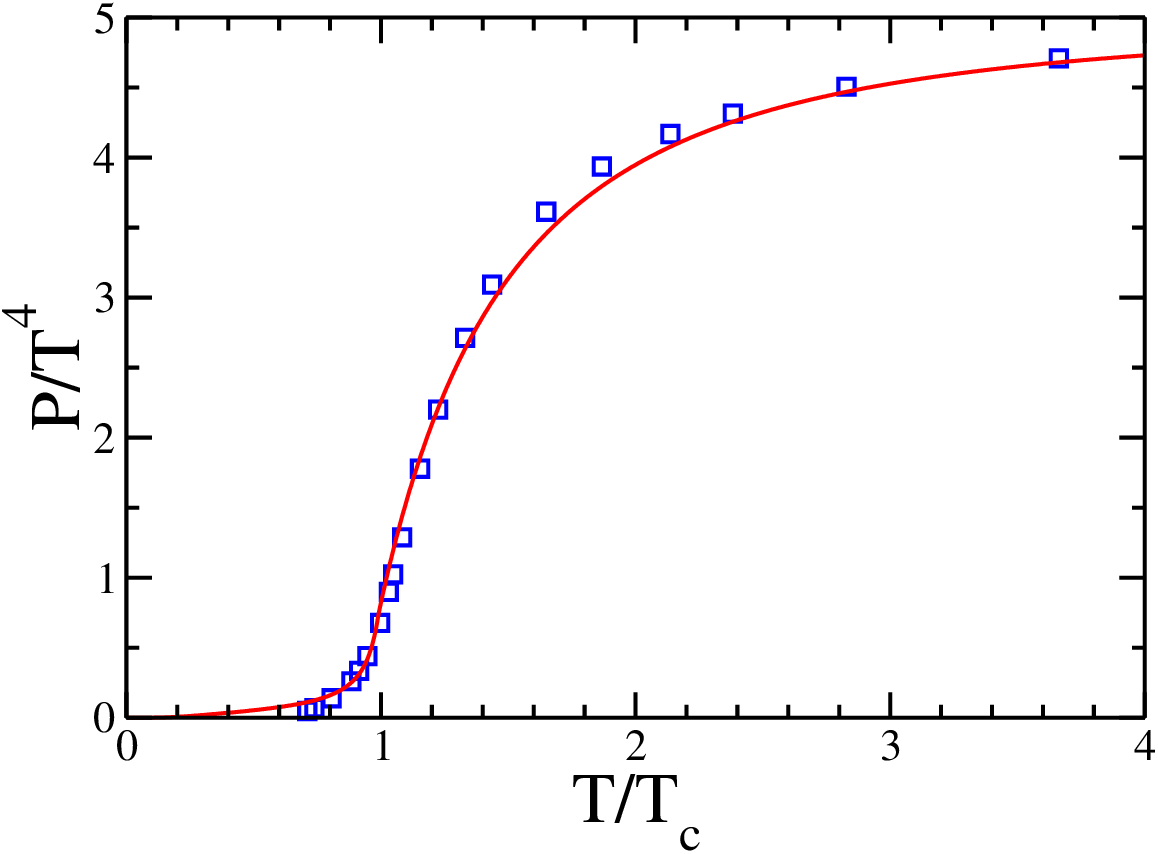,width=1.\textwidth}
\caption{(Color online) Pressure $P$ as a function of the temperature divided by
$T^4$ for $\tilde g=1.30$, $T_{\rm H}=205$ MeV and $a=9.17$ ${\rm MeV}^{1/2}$
in comparison to the lattice data from Ref.~\cite{Cheng:2007jq}.\label{Fig:PT4Nf03Tc196fit}}
\end{minipage}
\begin{minipage}{0.04\textwidth}
\hfill %
\end{minipage}%
\end{figure}

In Fig.~\ref{Fig:PT4Nf03Tc196fit} we show the lattice results of
the scaled pressure $P/T^4$ as a function of the temperature
expressed in units of the critical temperature $T_{\rm c}=196$ MeV
and the pressure from the virial expansion with the parameters
fixed above.
\begin{figure}[t]
\begin{minipage}[t]{0.45\textwidth}
\epsfig{figure=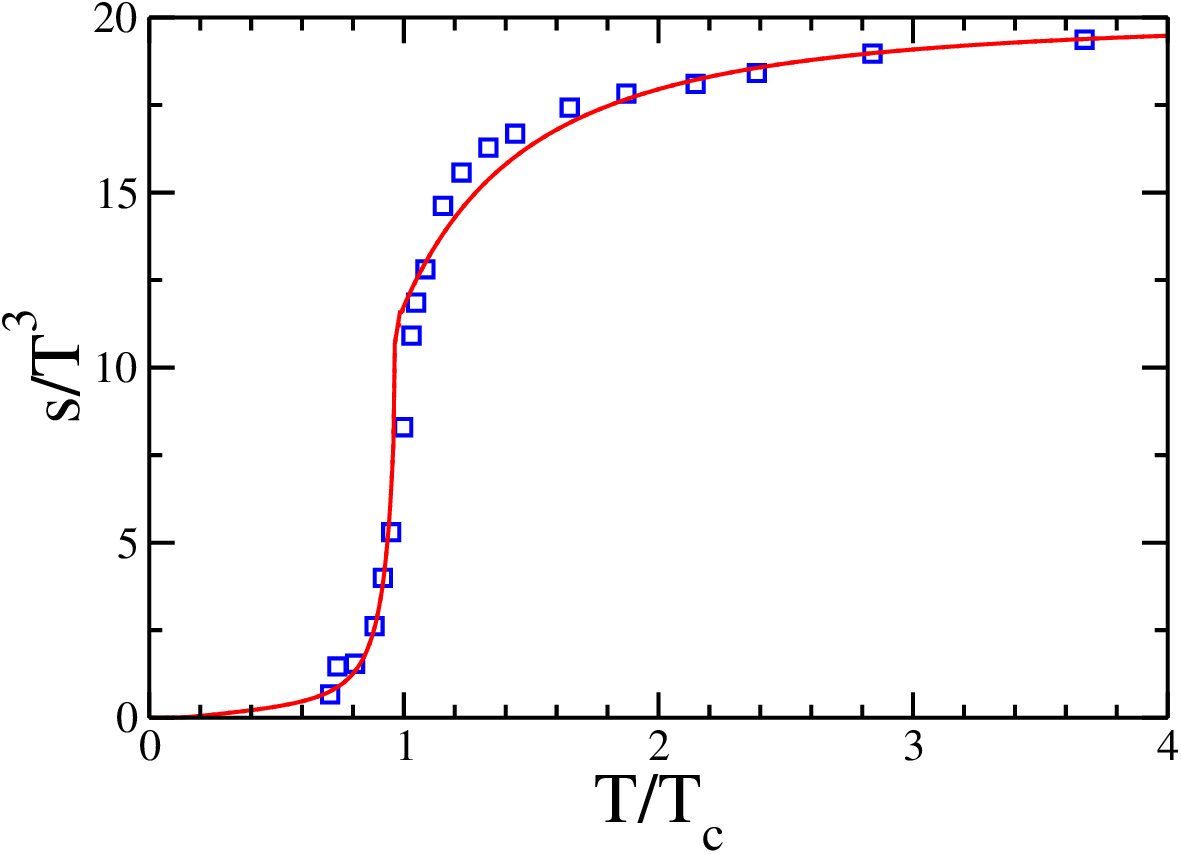,width=1.\textwidth}
\caption{(Color online) Entropy density $s$ as a function of the temperature
divided by $T^3$ for $\tilde g=1.30$, $T_{\rm H}=205$ MeV and $a=9.17$ ${\rm MeV}^{1/2}$ in
comparison to the lattice data from Ref.~\cite{Cheng:2007jq}.\label{Fig:ST3Nf03Tc196fit}}
\end{minipage}
\end{figure}

Since all thermodynamic quantities are connected by known
relations (cf. (\ref{eqn:Pdens})-(\ref{eqn:c^2_s})) we expect that
the lattice data for all quantities are compatible with the
results from our model, too. To confirm this we show in
Fig.~\ref{Fig:ST3Nf03Tc196fit} the entropy density $s$ divided by
$T^3$ as a function of the temperature expressed in units of the
critical temperature using Eq.(\ref{eqn:S}).

\begin{figure}[t]
\begin{minipage}{0.45\textwidth}
\epsfig{figure=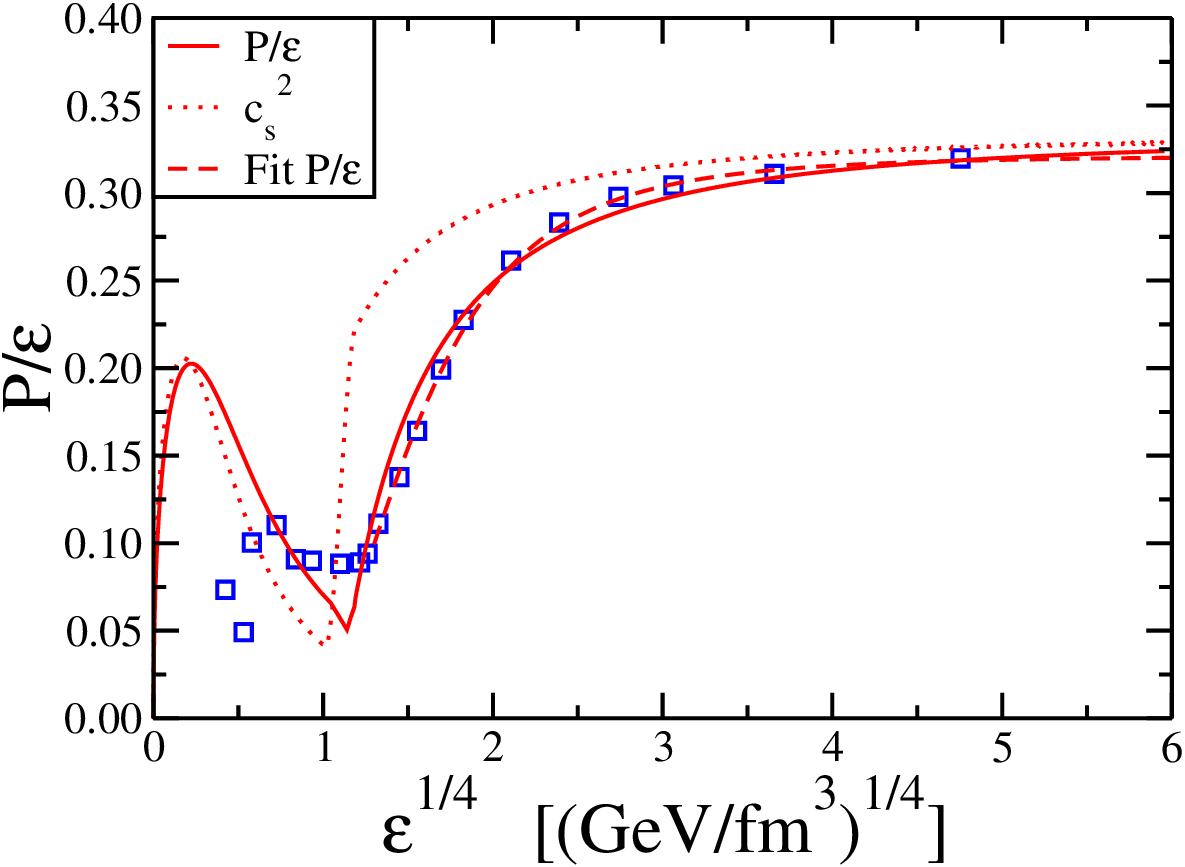,width=1.\textwidth}
\caption{(Color online) Ratio $P/\varepsilon$ as a function of the fourth root of
the energy density  for the critical temperature
$T_{\rm c}=196$ MeV in comparison to the lattice data from Ref.~\cite{Cheng:2007jq}.
\label{Fig:PEENf03Tc196fit}}
\end{minipage}
\end{figure}

Another interesting quantity is the square of the velocity of
sound $c^2_s$ defined by Eq.(\ref{eqn:c^2_s}). In the high
temperature limit as well as in the transition region, where the
derivative $d(P/\varepsilon)/d\varepsilon$ vanishes, $c^2_{\rm s}$
is given by $p/\varepsilon$. The results of our model for this
quantity (as a function of the fourth root of the energy density)
are shown in Fig.~\ref{Fig:PEENf03Tc196fit} for $T_{\rm c}=196$
MeV (solid)  in comparison with the lattice data (blue squares).
They show (as expected) a good agreement with the lattice data for
$\varepsilon^{1/4}/({\rm GeV/fm^3})^{1/4}\geq 1.3$ which is
equivalent to $T\geq T_{\rm c}$. The dashed line is the
parametrization in the high temperature region with the simple
{\em Ansatz}~\cite{Ejiri:2005uv}
\begin{equation}
\frac{P}{\varepsilon}=\frac{1}{3}\left(C-\frac{A}{1+B\varepsilon\;{\rm fm^3/GeV}}\right),
\end{equation}
that provides a good fit to the data in the interval
$1.3\leq\varepsilon^{1/4}/({\rm GeV/fm^3})^{1/4}\leq 6$ with
$C=0.964(5)$, $A=1.16(6)$ and $B=0.26(3)$~\cite{Cheng:2007jq}. The
current resolution and accuracy of lattice calculations for the
hadronic region clearly are not yet sufficient to allow for a
detailed comparison with any model calculation. The dotted line in
Fig.~\ref{Fig:PEENf03Tc196fit} indicates the exact square of the
sound velocity. The difference between this quantity and the ratio
$P/\varepsilon$ indicates the deviation from a linear dependence
between pressure and energy density valid at high temperature.
This linear behavior is confirmed in
Fig.~\ref{Fig:PEENf03Tc196fit} for high energy densities, where
$c_{\rm s}$ as well as $P/\varepsilon$ show the Stefan Boltzmann
limit of $1/3$.

\begin{figure}[t]
\begin{minipage}{0.45\textwidth}
\epsfig{figure=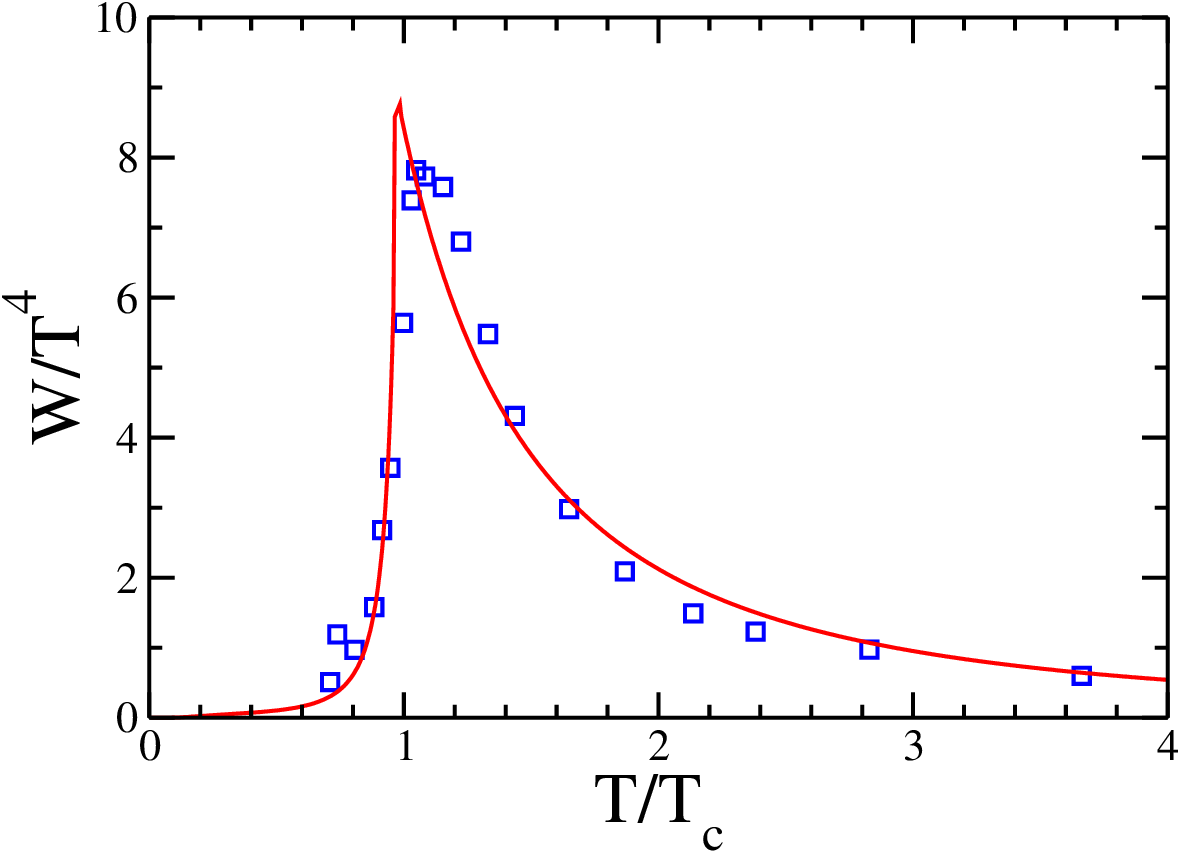,width=1.\textwidth}
\caption{(Color online) The interaction measure $W$ as a function of the
temperature divided by $T^4$ for $\tilde g=1.30$, $T_{\rm H}=205$
MeV and $a=9.17$ ${\rm MeV}^{1/2}$ in comparison to the lattice
data from Ref.~\cite{Cheng:2007jq}. \label{Fig:TraTNf03Tc196fit}}
\end{minipage}
\end{figure}

In the framework of the one-gluon exchange model the interaction
measure $W=\varepsilon-3P$ plays an important role and was
recently investigated in detail in Refs.~\cite{Megias:2008rm,Megias:2008dv}.
From the fit of $W$ in the deconfined region in gluon-dynamics the value of the dimension two
gluon condensate has been calculated in an independent way without
considering the Polyakov loop. However, the results of the two
approaches show only little deviation when using $1.26\leq\tilde g\leq 1.45$.
The calculation of the interaction measure using the
virial expansion starting with a model motivated by the fit to the
Polyakov loop may directly indicate a dynamical connection between
the two methods. For this reason we show in
Fig.~\ref{Fig:TraTNf03Tc196fit} the interaction measure divided by
the fourth power of the temperature $W/T^4$ as a function of the
temperature expressed in units of the critical temperature
$T_{\rm c}=196$ MeV, where the expected agreement with the lattice data
(blue squares) is observed.

\section{Finite quark chemical potential}\label{Sec:mu}


The extension of the virial expansion to non-vanishing quark
chemical potentials $\mu_q$ requires an explicit dependence of the
interaction on the quark chemical potential. Since  the
interaction is primarily determinated by the Debye mass $M$, we
may generalize the interaction for non-vanishing $\mu_q$ by
introducing an explicit dependence  for the Debye mass only, i.e.
$M(T)\longrightarrow M(T,\mu_q)$.\\ We parameterize this
dependence by
\begin{equation}
M(T,\mu_q)=M^{(0)}(T)F\left(\frac{\mu_q}{T}\right),
\end{equation}
where $M^{(0)}(T)=M(T,\mu_q=0)$ denotes the Debye mass at
vanishing chemical potential and $F$ describes the modification at
finite density. This assumption is equivalent to introduce a
$\mu_q$-dependence of the coupling $\tilde g$ (in our model),
\begin{equation}\label{def:g*}
\tilde g\longrightarrow\tilde g^*\left(\frac{\mu_q}{T}\right)=\tilde g\;F\left(\frac{\mu_q}{T}\right).
\end{equation}
Furthermore, perturbation theory suggests that the leading order
contribution to the correction is given by~\cite{Ejiri:2005mk}
\begin{equation}
F_{\rm PT}\left(\frac{\mu_q}{T}\right)=\sqrt{1+\frac{3N_{\rm f}}{(2N_{\rm c}+N_{\rm f})\pi^2}\left(\frac{\mu_q}{T}\right)^2}.
\end{equation}
For our calculations we generalize this expression as follows,
\begin{equation}
F\left(\frac{\mu_q}{T}\right)=A\left(\frac{\mu_q}{T}\right)F_{\rm PT}\left(\frac{\mu_q}{T}\right),
\end{equation}
with
\begin{equation}
A\left(\frac{\mu_q}{T}\right)=1+\frac{1}{\pi^2}\left(\frac{\mu_q}{T}\right)^2,
\end{equation}
where the value of the coefficient $\pi^{-2}$ is consistent for
$N_{\rm f}=3$ with Ref.~\cite{Peshier:2004ya}. For the hadronic
phase we use an effective modification of the constant $a$ given
formally in Eq.(\ref{def:g*}). Choosing in this case the
perturbative correction we obtain
\begin{equation}\label{def:a*}
a\longrightarrow a^*\left(\frac{\mu_B}{T}\right)=a \;F_{\rm PT}\left(\frac{\mu_B}{T}\right),
\end{equation}
where $\mu_B=3\mu_q$ indicates the baryonic chemical potential.\\

\begin{figure}[b]
\begin{minipage}{0.45\textwidth}
\includegraphics[width=\textwidth]{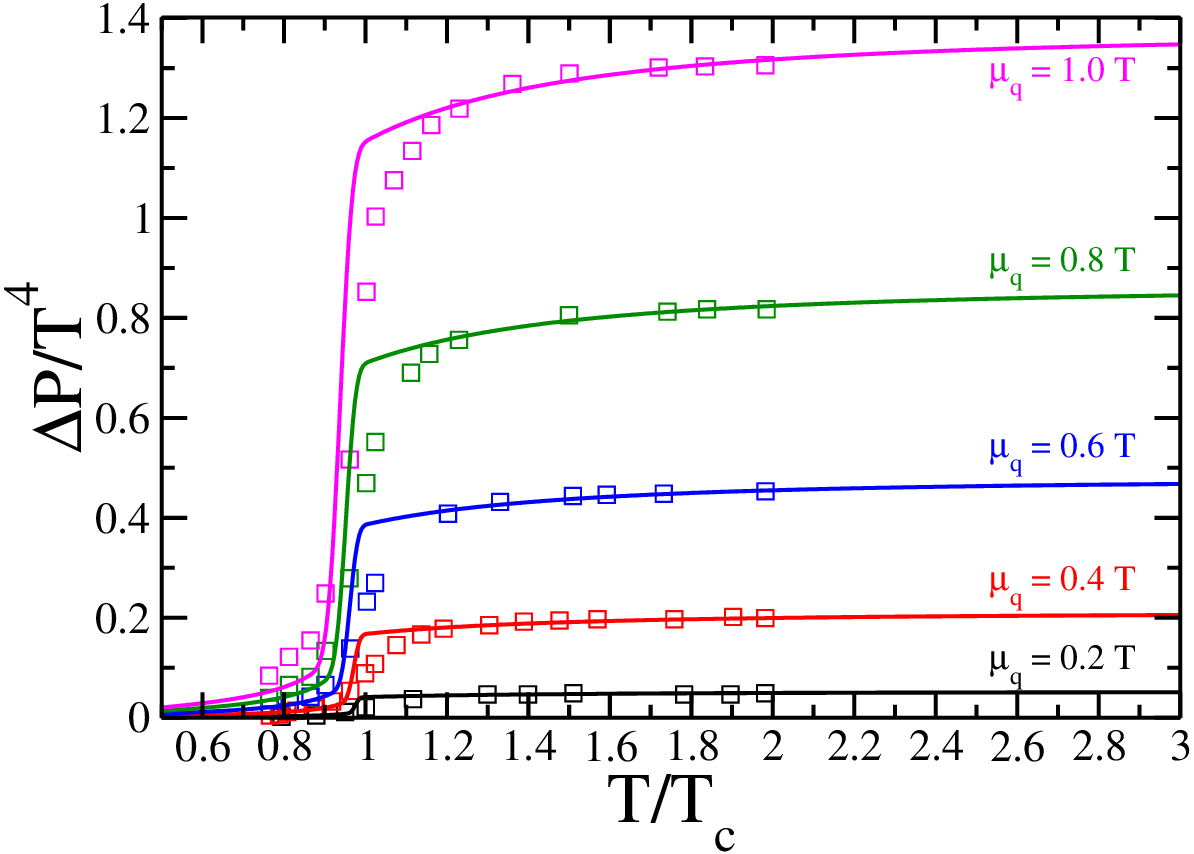}
\caption{(Color online) Scaled pressure difference as a function of the temperature for different values of $\mu_q/T$ compared to lattice calculations taken from Ref.~\cite{Allton:2003vx}.
\label{Fig:DPalphaNf03Tc196fit}}
\end{minipage}
\end{figure}

\begin{figure}[t]
\begin{minipage}[t]{0.45\textwidth}
\includegraphics[width=\textwidth]{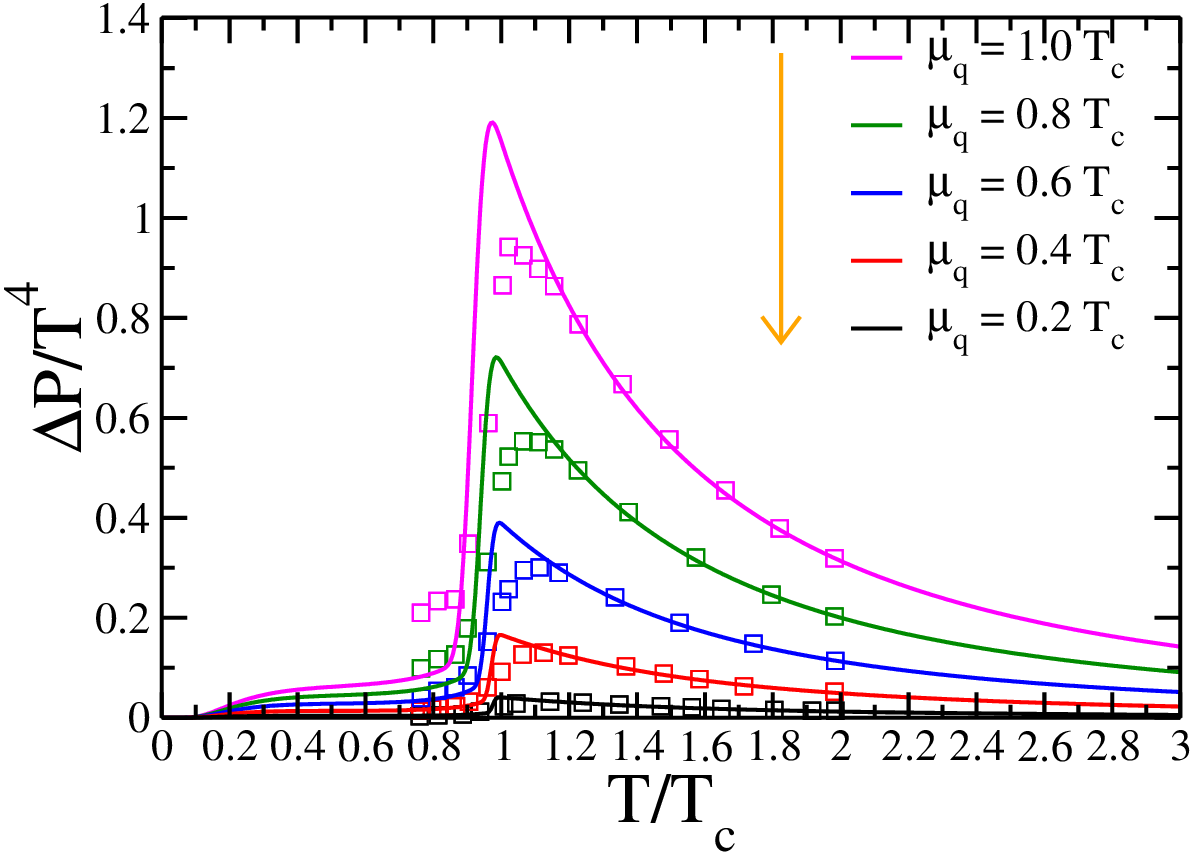}
\caption{(Color online) Scaled pressure difference as a function of the temperature for different values of $\mu_q$ compared to lattice calculations taken from Ref.~\cite{Allton:2003vx}.
\label{Fig:DPmuNf03Tc196fit}}
\end{minipage}
\end{figure}

\begin{figure}[b]
\begin{minipage}[t]{0.45\textwidth}
\includegraphics[width=\textwidth]{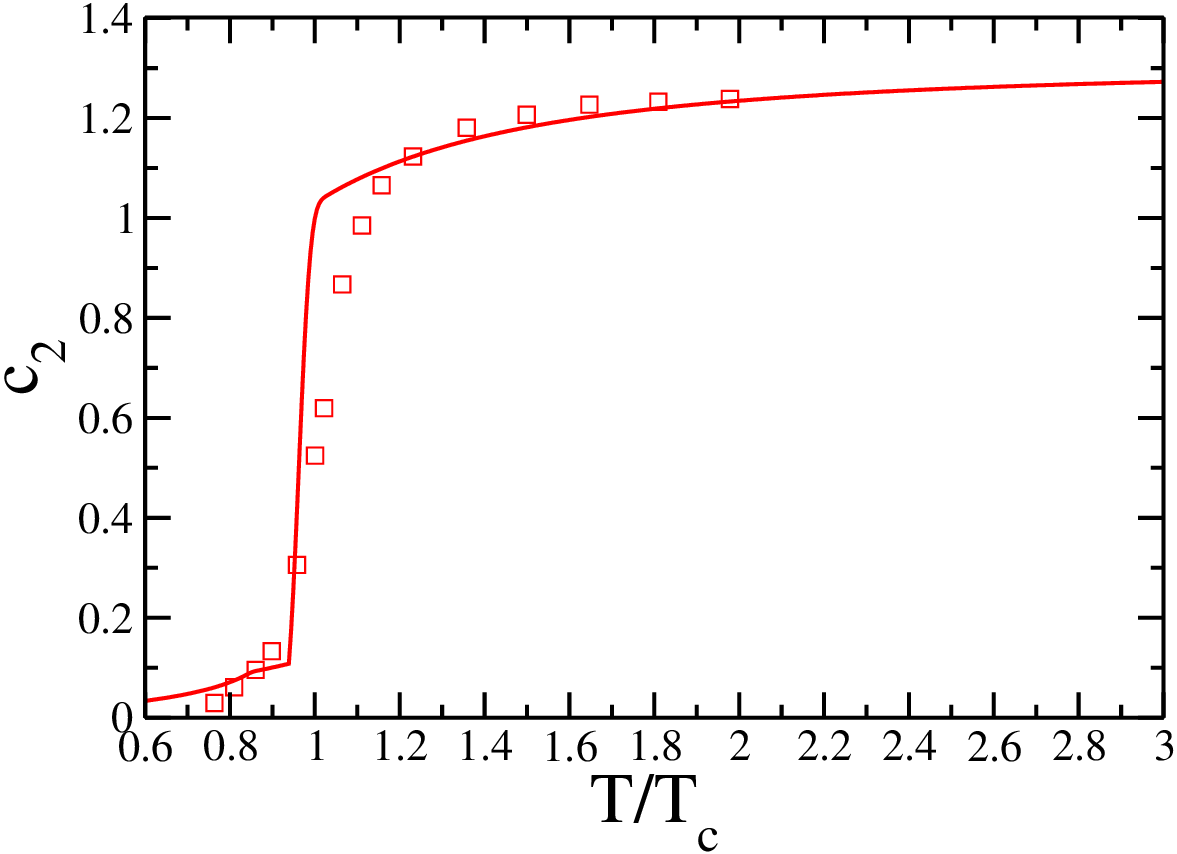}
\caption{(Color online) Coefficient $c_2(T)$ of the Taylor expansion for the pressure as function of $T/T_{\rm c}$ compared to lattice calculations taken from Ref.~\cite{Allton:2003vx}.}
\end{minipage}
\end{figure}

At nonzero chemical potential, typical quantities of interest that become accessible from lattice calculations are the (scaled) pressure difference defined as
\begin{equation}\label{def:DP}
\frac{\Delta P(T,\mu_q)}{T^4}=\frac{P(T,\mu_q)-P(T,\mu_q=0)}{T^4}
\end{equation}
and the scaled quark number density given by
\begin{equation}\label{def:nq}
\frac{n_q(T,\mu_q)}{T^3}=\frac{1}{T^3}\frac{\partial P(T,\mu_q)}{\partial\mu_q}=\frac{1}{T^4}\frac{\partial P(T,\mu_q)}{\partial(\mu_q/T)}.
\end{equation}
Since for $\mu_q\neq0$ direct Monte Carlo calculations are not applicable, higher order derivatives of the pressure with respect to $\mu_q/T$ at $mu_q=0$ are calculated and then the pressure is estimated using a Taylor expansion,
\begin{equation}\label{eqn:P-Taylor}
\frac{P(T,\mu_q)}{T^4}=\frac{\ln{Z(T,\mu_q)}}{T^3}=\sum_{n=0}^{\infty}c_n(T)\left(\frac{\mu_q}{T}\right)^n,
\end{equation}
where the expansion coefficients are given in terms of derivatives of $\ln{Z(T,\mu_q)}$, i.e.
\begin{equation}\label{def:c_n}
c_n(T)=\frac{1}{n!T^3}\frac{\partial^n\ln{Z}}{\partial(\mu_q/T)^n}.
\end{equation}
The series (\ref{eqn:P-Taylor}) is even in $\mu_q/T$ which
reflects the invariance of $Z(T,\mu_q)$ under exchange of
particles and antiparticles. The coefficients $c_2$ and $c_4$ have
been calculated for two-flavor QCD in Ref.~\cite{Allton:2003vx};
an extension to the sixth order is given in
Ref.~\cite{Allton:2005gk}. To evaluate the pressure and the other
thermodynamics quantities within the virial expansion we can
formally use the expansion given in Eq. (\ref{eqn:P}), where the
expansion parameter $\zeta$ now depends on the quark chemical
potential, i.e. $\zeta\longrightarrow \tilde\zeta(\mu_q)$. Using
the Taylor expansion of the pressure in the Stefan Boltzmann limit
we can explicitly give this $\mu_q$-dependence of $\zeta$. In this
case the coefficients are well known as
\begin{equation}\label{eqn:c_n^SB}
c^{\rm SB}_2=\frac{N_{\rm f}}{2},\qquad c^{\rm SB}_4=\frac{c_2}{(2\pi)^2},\qquad c ^{\rm SB}_n=0\quad{\rm for}\quad n\geq 6.
\end{equation}
Now we can rewrite the generalized expansion parameter as
\begin{equation}\label{eqn:zetamu}
\tilde\zeta(\mu_q)=\zeta+c^{\rm SB}_2T^2\mu_q^2+c^{\rm SB}_4\frac{\mu_q^4}{T}.
\end{equation}
The results for the scaled pressure difference as a function of
the temperature (expressed in units of the critical temperature
$T_{\rm c}=196$ MeV) are shown in
Fig.~\ref{Fig:DPalphaNf03Tc196fit} for various values of
$\mu_q/T$. The data points are the lattice results for $N_{\rm f}=2$
from Ref.~\cite{Allton:2003vx} scaled to $N_{\rm f}=3$.
Lattice results for three-flavor QCD are not yet available and
partial data for $N_{\rm f}=2+1$ are preliminary
only~\cite{Hegde:2008rm,Karsch:2008fe}. The agreement between our
results and the lattice simulations is quite satisfactory. In the
region around $T_{\rm c}$ there is a deviation from the scaled
lattice points.

This is better visible in an analogue representation of the results
in Fig.~\ref{Fig:DPmuNf03Tc196fit},
where the scaled pressure difference is shown as a function of the temperature
expressed in units of the critical temperature $T_{\rm c}=196$ MeV for
various values of the quark chemical potential $\mu_q$. We note an
overestimation of the pressure difference  around the critical temperature.
This can be understood by considering the possible errors of the scaling
procedure as well as by the errors given by the (truncated) Taylor expansion itself.

\begin{figure}[t]
\begin{minipage}[t]{0.45\textwidth}
\includegraphics[width=\textwidth]{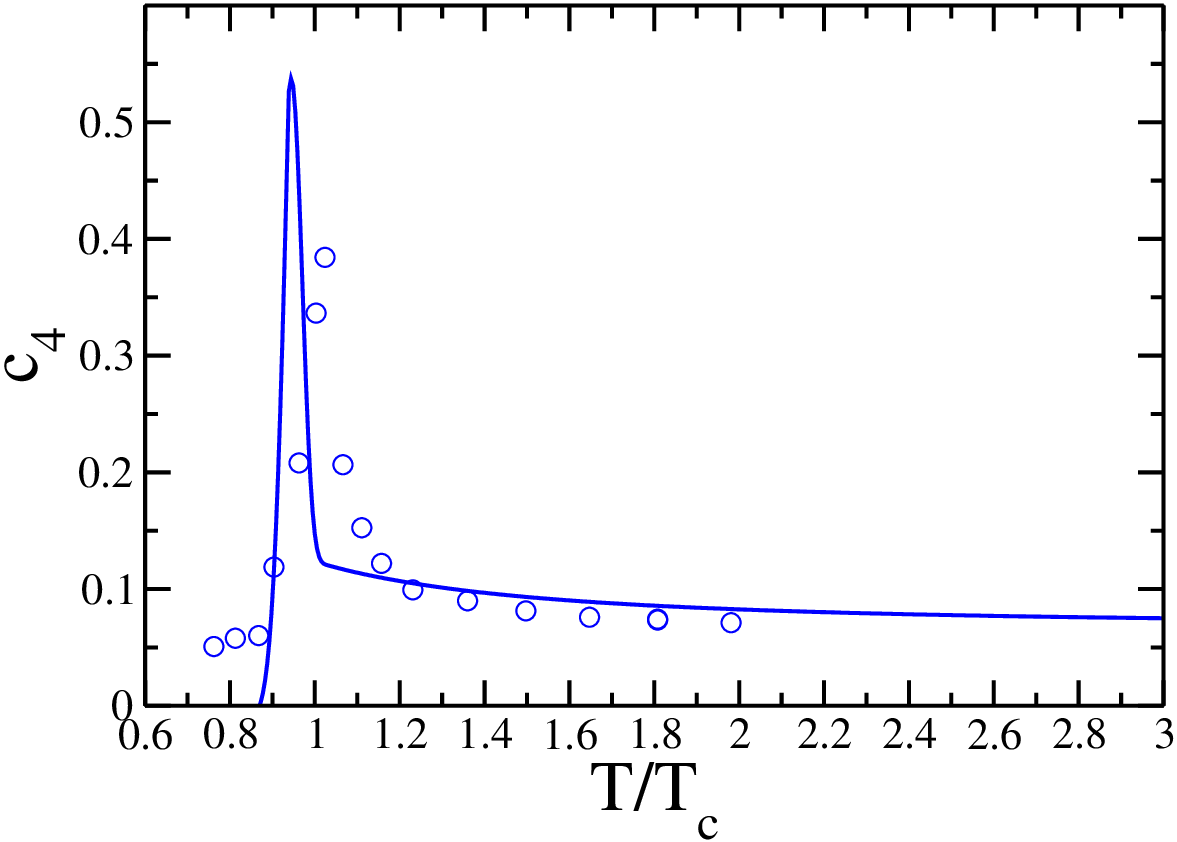}
\caption{(Color online) Coefficient $c_4(T)$ of the Taylor expansion for the pressure as a
function of $T/T_{\rm c}$ compared to lattice calculations taken
from Ref.~\cite{Allton:2003vx}.
\label{Fig:c2c4Nf03Tc196fit}}
\end{minipage}
\end{figure}

\begin{figure}[b]
\begin{minipage}[t]{0.45\textwidth}
\includegraphics[width=\textwidth]{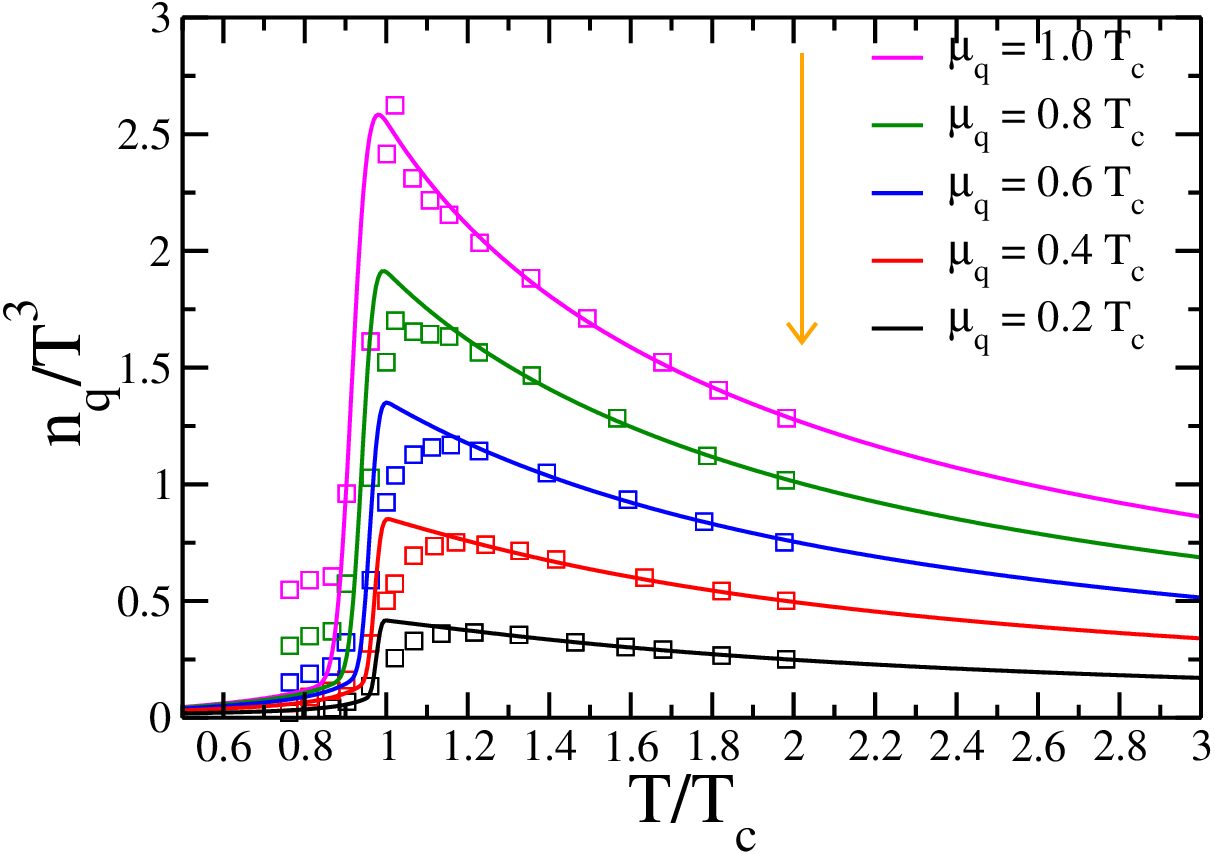}
\caption{(Color online) Scaled quark number density as a function of the temperature for
different values of $\mu_q$ compared to lattice calculations taken from Ref.~\cite{Allton:2003vx}.
\label{Fig:qdensmuNf03Tc196fit}}
\end{minipage}
\end{figure}

We can now calculate the coefficients $c_2(T)$ and $c_4(T)$ of the
Taylor expansion for the pressure by the least-square method. The
comparison with the scaled lattice results from Ref.
~\cite{Allton:2003vx}, shown in Fig.~\ref{Fig:c2c4Nf03Tc196fit},
is quite satisfactory. Analogue comparisons with quasiparticle
models are given in Refs.~\cite{Bluhm:2004qr,Bluhm:2007cp}.
Although these coefficients allow to directly evaluate the scaled
quark number density, we prefer to use the definition given in
Eq.(\ref{def:nq}) because the Taylor sum is only an expansion up
to the fourth order and its validity for strongly interacting
system is not so clear (cf. Ref. ~\cite{Buballa:2008ru}). Our
results for $n_q/T^3$ as a function of the temperature (in units
of the critical temperature $T_{\rm c}=196$ MeV)  are shown in
Fig.~\ref{Fig:qdensmuNf03Tc196fit} for various values of the quark
chemical potential $\mu_q$. Also in this case, the comparison
between our virial calculation and the corresponding (scaled)
lattice data shows a slight overestimation close to $T_c$.

\begin{figure}[t]
\begin{minipage}[b]{0.45\textwidth}
\includegraphics[width=\textwidth]{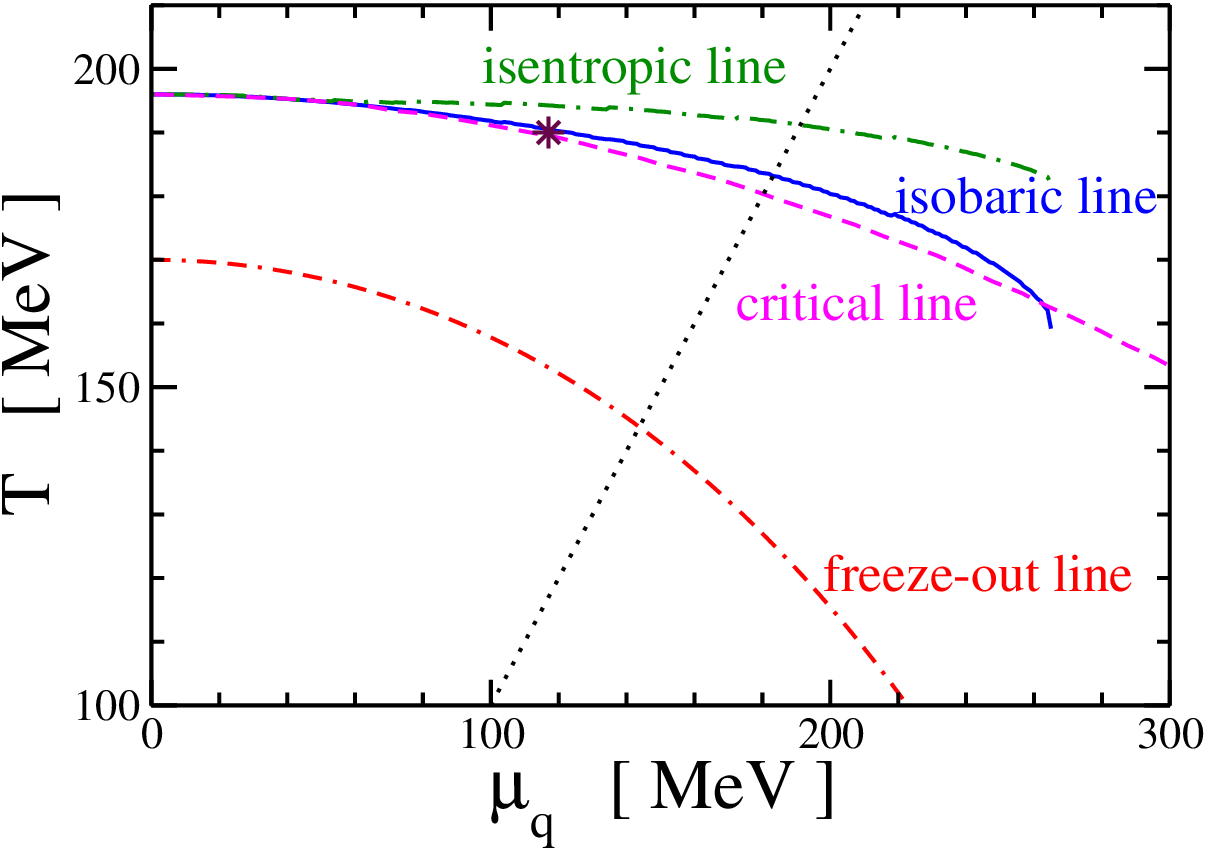}
\caption{(Color online) Overview of the QCD phase diagram with isobaric (solid)
and isentropic lines (dash-dot) in comparison to the phase
transition line (dashed) estimated in Ref.~\cite{Allton:2002zi}.
Also shown is the hadron chemical freeze-out line
(dot-dash-dashed) in the parametrization given in
Ref.~\cite{Karsch:2006xs}. The dotted line ($T=\mu_q$)
demonstrates the expected range of validity for the lQCD
calculations. \label{Fig:phase}}
\end{minipage}
\end{figure}

Using the results for the pressure and the entropy density (from
Eq.(\ref{eqn:S})) we can directly investigate the structure of the
QCD phase diagram in the $\mu_q-T$ plane. Starting from the point at
zero chemical potential, i.e. $(0,T_{\rm c})$, we calculate the
isobaric ($P= const.$) as well as isentropic ($s= const.$) lines
in the phase diagram which are shown in Fig~\ref{Fig:phase} by the
solid and dot-dashed lines, respectively. The resulting isentropic
temperature at the phase boundary is practically constant up
$\mu_q \approx 200$ MeV, whereas the isobaric temperature at the
phase boundary (solid line) drops with increasing chemical
potential. This behavior quantitatively agrees with the critical
line (dashed) from lattice calculations within the reweighting
method from Ref.~\cite{Allton:2002zi}. The observation that the
critical phase transition is a line at constant pressure is also
confirmed by lattice simulations in Ref.~\cite{Allton:2002zi}. The
dotted line in Fig. 10 shows the region $\mu\leq T$, where the
lattice results are expected to be valid. For comparison we
display also the hadron chemical freeze-out line (dash-dash-dot)
using the parametrization given in Ref.~\cite{Karsch:2006xs}. Note
that the hadron freeze-out - determined experimentally from hadron
ratios -  clearly occurs in the hadronic phase for all $\mu_q$
with a sizeable gap in temperature relative to the phase
transition line.

Furthermore, a detailed investigation of the isobaric line leads
to a first estimate for the critical endpoint $(\mu_E,T_E)$, where
the crossover in the low density region changes to a first order
phase transition. Following Ref.~\cite{Fodor:2004nz} the curvature
of the crossover - separating the QGP and the hadronic phase - is
given by
\begin{equation}
T/T_c=1-\tilde C\mu_q^2/T_c^2.
\end{equation}
To evaluate the critical endpoint we perform  a corresponding
parametrization of our calculated isobaric line. This only works
if $\chi^2\leq 1$, i.e.  a maximun of the chemical potential
exists which satisfies this condition; the latter value defines the
chemical potential of the endpoint. From our analysis we obtain
$\mu_E=117\pm 5$, which is in agreement with the results of the
lattice calculation $\mu_E^{(Lattice)}=120\pm10$ from
Ref.~\cite{Fodor:2004nz}. The corresponding critical point
$(\mu_E,T_E)=(117,190)$ MeV is shown as a star in
Fig~\ref{Fig:phase}.

\begin{figure}[t]
\begin{minipage}[b]{0.45\textwidth}
\includegraphics[width=\textwidth]{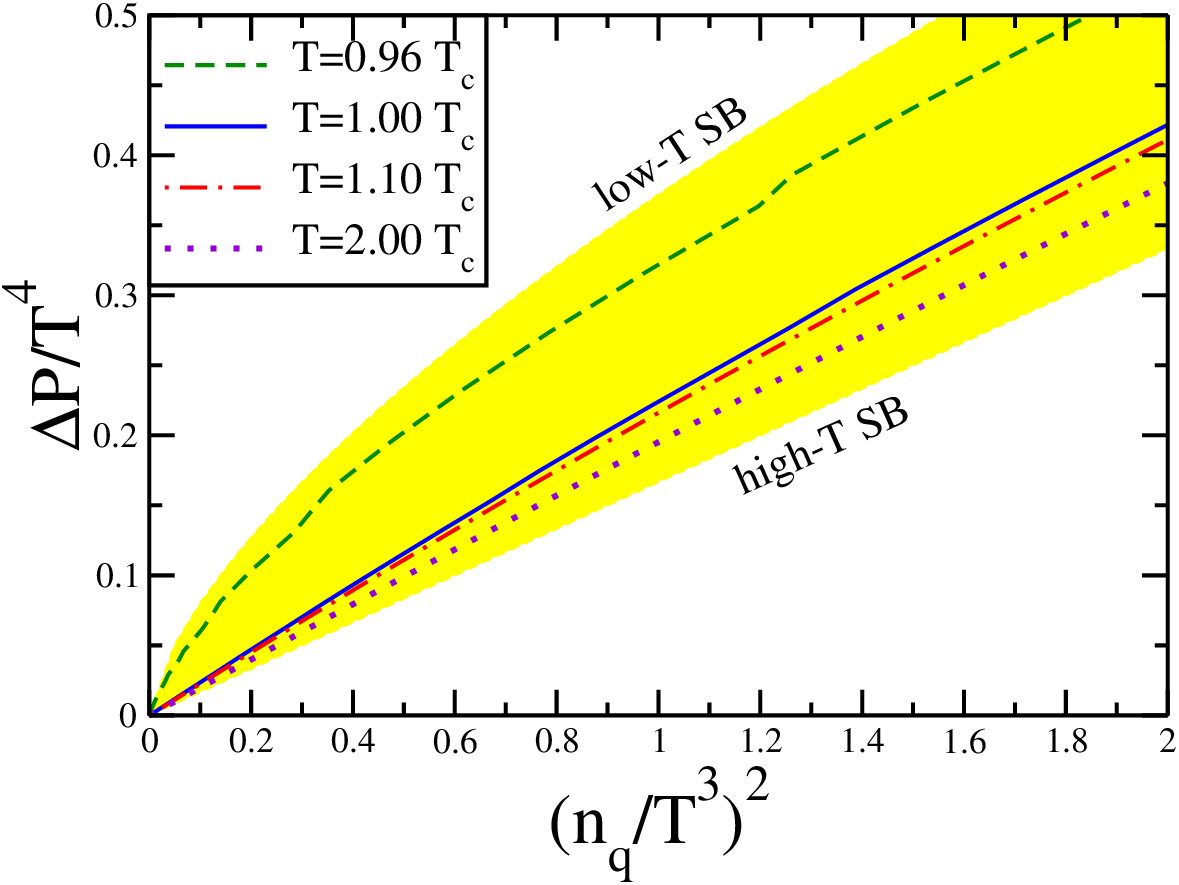}
\caption{(Color online) Equation of state $\Delta(p/T^4)$ vs. $(n_q/T^3)^2$ for
various temperatures. The continuum Stefan Boltzmann forms for low
and high $T$ (given by (\ref{SB1}) and (\ref{SB2})) are also shown
as functions of $T/T_{\rm c}$. \label{Fig:EoS}}
\end{minipage}
\end{figure}

We recall that the main aim of this paper is to evaluate the QCD
equation of state in terms of physically measurable quantities.
Following Ref.~\cite{Allton:2003vx} we calculate the equation of
state $\Delta(p/T^4)$ vs $(n_q/T^3)^2$ for different temperatures.
The results are shown in Fig.~\ref{Fig:EoS}, where the yellow
region is delimited by the low-$T$ SB (upper boundary) and by the
high-$T$ SB limit (lower boundary). For temperatures near- but
below - $T_{\rm c}$ the equation of state shows a form resembling
the low-$T$ SB limit
\begin{equation}\label{SB1}
\Delta P=\frac{1}{4}(\pi^2/N_{\rm f})^{1/3}n_{\rm q}^4 .
\end{equation}
As $T$ increases though the critical temperature, the EoS changes
from this functional dependence to the stiffer form of the
high-$T$ SB limit given by
\begin{equation}\label{SB2}
\Delta
P=\frac{1}{2N_{\rm f}}n_{\rm q}^2 .
\end{equation}
The quadratic coefficient depends smoothly on the temperature and therefore
shows a sizeable deviation from the exact Stefan Boltzmann limit
also for $T=T_{\rm c}$.

\section{Conclusions}\label{Sec:Concl}

We have performed an investigation of QCD thermodynamic quantities
within a virial expansion approach. While hadronic matter is
described by a generalized resonance-gas model, we fix the
interaction in the QGP phase by lattice calculations for the heavy
quark free energy. We adopt a phenomenological model which
includes non-perturbative effects from dimension two gluon
condensates that describe the free energy of quenched QCD very
well at vanishing chemical potential. Explicit expressions for the
partition function and the pressure in the deconfined phase are
derived. Our results for $\mu_q=0$ and $N_{\rm f}=3$ for pressure,
entropy density, speed of sound and interaction measure at nonzero
temperature  compare well with three-flavor QCD lattice
calculations with almost physical masses from
Ref.~\cite{Cheng:2007jq}. We find that a coupling parameter
$\tilde g=1.30$ provides a good agreement for all quantities. Note
that this choice is close to the value determined in Ref.
~\cite{Megias:2007pq} from a fit to lattice data for the Polyakov
loop in gluon-dynamics, i.e. $\tilde g_{\rm PL}=1.26$. The
difference in the coupling parameters most likely is due to the
different quark masses employed.

Furthermore, we have extended this approach to finite densities
introducing an explicit $\mu$-dependence of the interaction (via
the Debye mass following perturbation theory).  This allows to
calculate the thermodynamic quantities of interest such as
pressure and quark-number density etc. also at finite $\mu_q$; the
results again compare well with lattice calculations scaled from
$N_{\rm f}=2$ to $N_{\rm f}=3$ with minimal deviations at the
critical temperature $T_c$. These differences are most visible in
the pressure, become smaller for the scaled quark number density
and lead to the open question about the validity of the Taylor
expansion and its thermodynamic consistency.

Additionally, we have investigated the structure of the phase
diagram also in the region inaccessible to lattice calculations,
i.e. $\mu_q\geq T$. We have calculated the isentropic and
specially the isobaric line, which agrees remarkably well with the
cross over transition evaluated by lattice calculations with the
reweighting method. The direct knowledge of the partition function
allows to calculate the equation of state of the quark gluon
plasma which also reproduces very well the predictions of the
lattice calculations. The EoS clearly shows two decisive
properties of the QCD phase diagram: {\em i)} There is an
unequivocal transition from the hadronic matter to deconfined
matter as is evident from the change of the functional form of the
equation state from the typical Stefan Boltzmann behavior for low
to high temperatures. {\em ii)} The QGP is a strongly correlated
system and sensitively differs from the ideal gas limit. Also at
$T=2T_{\rm c}$ there is a remarkable deviation of the EoS of the
QGP from the Stefan Boltzmann limit which is a clear signal of
surviving correlations in the deconfined phase.

Our equation of state has been dynamically calculated with a
particular form of the interaction. Using the same interaction in
molecular dynamical calculations we may systematically investigate
relevant properties of the QGP such as the pair correlation
function $g(r)$, the plasma parameter $\Gamma$, the structure
factor $S(q)$ or the shear viscosity $\eta$. Further interesting
applications   in
this context (cf. Refs.~\cite{Mrowczynski:2007hb,Gelman:2006xw})
are relativistic molecular dynamic simulations, where
the phenomenological quark-quark interaction used here can be
implemented as well as the QCD equation of state. This will allow
to study partonic systems also out of equilibrium.

{\em Acknowledgements:} Work supported by the Deutsche
Forschungsgemeinschaft (DFG).

\end{document}